\documentclass[sigconf]{acmart}
\usepackage{graphicx}
\usepackage{tipa}
\usepackage{lscape}
\usepackage{xcolor}
\usepackage{lmodern}
\usepackage{eurosym}
\usepackage{url}
\usepackage{tabularx} 
\usepackage[caption = false]{subfig}
\usepackage{tabularx}
\usepackage{multirow}
\usepackage{hhline}
\usepackage{tikz}
\usetikzlibrary{arrows.meta,positioning,shapes,decorations.pathreplacing}

\newcolumntype{L}{@{}l@{}} 

\copyrightyear{2025} 
\acmYear{2025} 
\setcopyright{rightsretained} 

\acmConference[EASE 2025]{The 29th International Conference on Evaluation and Assessment in Software Engineering}{17--20 June, 2025}{Istanbul, Türkiye}

\acmDOI{10.1145/xxx.xxxx}
\acmISBN{978-1-4503-xxx/xxx}

\setcopyright{none}
\renewcommand\footnotetextcopyrightpermission[1]{}

\begin{document}

\title{Cross-Functional AI Task Forces (X-FAITs) for AI Transformation of Software Organizations}


\author{Lucas Gren}
\affiliation{%
  \institution{Chalmers University of Technology and\\ University of Gothenburg}
    \city{Gothenburg}
  \country{Sweden}}
\email{lucas.gren@lucasgren.com}

\author{Robert Feldt}
\affiliation{%
  \institution{Chalmers University of Technology and\\ University of Gothenburg}
    \city{Gothenburg}
  \country{Sweden}}
\email{robert.feldt@chalmers.se}

\begin{abstract}

This experience report introduces the Cross-Functional AI Task Force (X-FAIT) framework to bridge the gap between strategic AI ambitions and operational execution within software-intensive organizations. Drawing from an Action Research case study at a global Swedish enterprise, we identify and address critical barriers such as departmental fragmentation, regulatory constraints, and organizational inertia that can impede successful AI transformation. X-FAIT employs force field analysis, executive sponsorship, cross-functional integration, and systematic risk assessment strategies to coordinate efforts across organizational boundaries, facilitating knowledge sharing and ensuring AI initiatives align with objectives. The framework provides both theoretical insights into AI-driven organizational transformation and practical guidance for software organizations aiming to effectively integrate AI into their daily workflows and, longer-term, products.

\end{abstract}



\maketitle

\section{Introduction}
As software organizations increasingly embed artificial intelligence (AI) into their products and workflows, tensions frequently arise between strategic goals and everyday execution. Even before the recent explosion of interest in generative AI, projections showed most companies would adopt at least one form of AI technology by 2030.

However, evidence consistently shows that despite significant investments, many organizations struggle to reap the full benefits of AI initiatives \cite{gudigantala2023ai,holmstrom2025}. Often, this gap emerges from internal fragmentation, i.e.\ when departments work in silos, it becomes difficult to share expertise, resulting in duplicated efforts and inconsistent handling of risks across the organization.



This paper introduces the Cross-Functional AI Task Force (X-FAIT, pronounced \textipa{/Eks.feIt/}) framework to address the challenge of effectively coordinating AI efforts across multiple business functions. Drawing on insights from our case company's transformation, we show how a dedicated task force can bridge traditionally separate departments (such as Digital Solutions, Human Resources, Research and Development, and Global Sales) to accelerate AI adoption without compromising governance or risk management.

The task force serves not only as a mechanism for coordinating AI initiatives but also as a platform for cross-departmental learning and knowledge sharing. By systematically capturing evidence of successful AI projects and their organizational impact, the task force can inspire broader confidence and enthusiasm for AI, encouraging wider adoption across the company. Previous research has largely focused either on the technical details of AI implementation~\cite{wang2025automating} or high-level strategic planning~\cite{holmstrom2025}, neglecting the intermediate structures that effectively link these dimensions. Our framework addresses this gap, contributing to organizational transformation theory while providing practical guidance for software companies seeking to integrate AI seamlessly into their operations.

\section{Related Work}
\subsection{AI Transformation Research}
Recent research emphasizes the need for structured approaches to AI transformation. For example, \citet{holmstrom2025} propose a framework built around three pillars: automation, augmentation, and data richness. Automation involves delegating tasks fully to machines, augmentation enhances human capabilities, and data richness ensures high-quality data for AI systems. Other studies have examined narrower aspects, such as AI governance \cite{papagiannidis2023toward} and organizational readiness \cite{holmstrom2022ai}. However, even recent studies focus primarily on predictive machine learning techniques, overlooking the significant advances in generative AI, which now constitute a substantial portion of AI use cases.

Addressing this gap, recent research like \citet{tavantzis2024human} highlights critical organizational and psychological dimensions of AI transformation, including communication strategies, ethical considerations, readiness for AI adoption, resistance to change, workforce skills, and strategic deployment of AI. Although these areas require further exploration, organizations urgently need practical guidance to overcome these challenges and harness AI's full potential. This paper provides concrete recommendations and tools to achieve that goal.

\subsection{Risk Management in AI Implementation}
The AI-SEAL \cite{feldt2018ways} taxonomy presents a framework for understanding AI applications in software engineering through three carefully delineated dimensions. The first dimension, Point of Application (PA), categorizes AI use according to its position in the software lifecycle: at the \emph{process} level (supporting developers without changing the code itself), the \emph{product} level (directly modifying or generating source code), and the \emph{runtime} level (actively controlling system behavior as software runs). Moving from process to runtime involves increased risk and reduced opportunity for human oversight, similar to loosening the reins on an automated system. The second dimension, Type of AI Technology (TAI), emphasizes selecting the appropriate type of AI technology for each specific task. The choice should match the available data, task requirements, and organizational needs, significantly influencing the effectiveness and reliability of the AI solution. The AI-SEAL paper pre-dated much of the generative AI ``revolution'' and thus does not provide specific support or discuss it.

The third dimension, Level of Automation (LA), adapted from \citet{parasuraman2000model}, describes how independently the AI functions. It ranges from Level 1, where humans retain complete decision-making control, to Level 10, where AI makes decisions autonomously and informs humans only if it chooses. In our analysis, contemporary generative AI systems and Large Language Models usually operate between levels 3 (narrows the options down to a few) and 7 (executes automatically, then necessarily informs the human), representing moderate to significant autonomy and corresponding risks.

Together, these three dimensions create a structured way to analyze AI's potential benefits and associated risks within software engineering. Higher values across dimensions indicate greater transformative potential but also demand careful risk management.

\subsection{Organizational Change}

Lewin's force field analysis \cite{lewin1939field} remains foundational for understanding how organizations change. His model describes the organization's current state as a balance between two opposing forces: those pushing for change and those holding it back. To create effective change, leaders must shift this balance, either by amplifying supportive forces, weakening resistance, or, ideally, both.

By carefully mapping these forces, leaders guiding AI transformations can more clearly identify where to intervene. 
For a deeper exploration of organizational change methods in the context of AI transformation of software organizations, see \citet{tavantzis2024human}.

\section{Case}
The case company is a provider of advanced equipment, including digital components critical to its functionality. Founded in Sweden, it operates in more than 30 countries with 10,000+ employees, maintaining its market presence through innovation and sustainability. This company was selected for our iterative Action Research study due to its early interest in AI transformation and its proactive investment in AI-driven initiatives.

As part of a heavily regulated industry, the company’s strategies and development processes are shaped by compliance requirements and international standards. Its distributed organizational structure grants autonomy to regional and functional units while operating within a complex decision-making hierarchy. The regulatory landscape and structured decision-making processes provide a unique setting for studying AI adoption in large and compliance-focused enterprises.

While the AI transformation in this software-driven enterprise has been ongoing for some time, it remains in its earlier stages, requiring further refinement and learning. The insights presented here are drawn from direct engagement with the company and reflect the structured approach to AI adoption that has resulted from early action research iterations. Generalization to less regulated industries or smaller, less distributed organizations may be less straightforward, as such companies may face different constraints and transformation dynamics. In the following sections, we describe our framework and its application, and the early benefits observed in the case company.

\section{X-FAIT Framework \& Application}
The framework introduced in this paper (the Cross-functional AI Task Force)
is a dedicated team, the actual AI Task Force, and a set of practical, analytical tools. Together, they are designed to coordinate AI efforts across organizational boundaries, foster shared learning, and document AI’s positive impact to drive broader adoption. By integrating expertise from diverse organizational units, the task force builds internal support, strengthens buy-in, and enhances the overall effectiveness of AI initiatives. Additionally, it serves as a central knowledge hub, providing practical guidance to departments navigating AI implementation.

Two foundational elements underpin our framework: Lewin's classic model of organizational change \cite{lewin1939field} and the AI-SEAL taxonomy \cite{feldt2018ways}. Force field analysis, a key component of Lewin's model, played a dual role: first, it initially guided the formation and design of the task force by identifying the main barriers to AI adoption, and second, it continues to serve as a tool for refining the task force’s approach over time. By systematically assessing shifting forces that either hinder or drive AI adoption, the task force can adapt its strategies, overcome new challenges, and fine-tune its way of working.


The AI-SEAL taxonomy complements this by identifying where AI should be applied, selecting the most suitable AI technologies, and determining the appropriate level of autonomy based on business needs, available data, and risk tolerance. We have extended mainly its second dimension to enable more detailed risk assessments and improve applicability to recent generative AI technologies, but also extended and adapted its other dimensions. Additionally, we have expanded the framework’s portfolio perspective by incorporating diversity across its three dimensions, ensuring that AI adoption maximizes learning and long-term adaptability.

\subsection{X-FAIT Risk Assessment Integration}
Risk integration is achieved through two key activities: (1) Dimensional Risk Analysis and (2) Risk-Aware Implementation Sequencing.
\paragraph{\textbf{Dimensional Risk Analysis}}
We extend the AI-SEAL taxonomy to support task-level risk-reward assessments across three dimensions:

\textit{Point of Application (PA)}: The task force assesses whether the AI initiative affects the process, product, or runtime levels, with stricter risk management protocols for higher-level applications. We expand this dimension to include an analysis of the targeted ``pain point''--whether the AI solution addresses cost or resource inefficiencies, automates repetitive tasks, or improves quality.

\textit{Type of AI Technology (TAI)}: Different AI technologies carry distinct risk profiles. We extend this dimension to account for the organization’s prior experience with the technology, its suitability for the specific task, and the availability of relevant data.

\textit{Level of Automation (LA)}: AI initiatives are categorized on an automation scale of 1--10. More stringent governance measures apply as automation levels increase, particularly beyond level 5 (executes the option if the human approves). This dimension also ensures that the automation levels align with the actual needs rather than aiming at unnecessary complexity or automation.

\paragraph{\textbf{Risk-Aware Implementation Sequencing}}
\textit{Low-Risk Pilot Initiatives:} Early implementations focus on process-level, low-automation tasks to build organizational expertise and demonstrate value with minimal risk.

\textit{Progressive Risk Tolerance:} As experience grows, the task force introduces higher-risk applications while incrementally strengthening governance mechanisms.

\textit{Portfolio Diversity:} AI adoption follows a balanced approach, selecting projects that vary across the risk dimensions above. This strategy maximizes organizational learning about new AI technologies, their application domains, and optimal automation levels.

This structured risk assessment distinguishes the X-FAIT framework from conventional IT transformation approaches, ensuring that AI implementation decisions systematically integrate risk considerations.

\subsection{Applying the Force Field Analysis in X-FAIT}
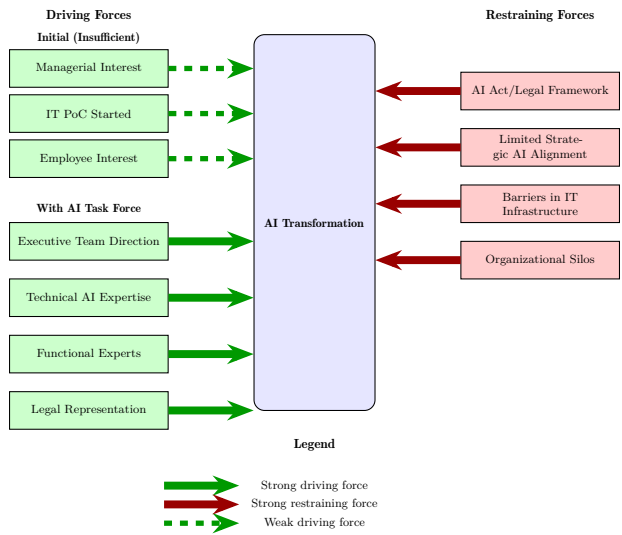
\begin{figure}
\centering
\begin{tikzpicture}[scale=0.50, transform shape]
    \tikzset{
        driving/.style={draw, fill=green!20, text width=4cm, minimum height=1cm, align=center},
        restraining/.style={draw, fill=red!20, text width=4cm, minimum height=1cm, align=center},
        arrow/.style={thick, -{Stealth[length=10pt]}, line width=3pt},
        dasharrow/.style={thick, dashed, -{Stealth[length=10pt]}, line width=2pt},
        title/.style={font=\bfseries\large},
        subtitle/.style={font=\bfseries}
    }
    
    \node[draw, rounded corners, fill=blue!10, text width=3cm, minimum height=10cm, align=center] (center) at (0,0) {\textbf{AI Transformation}};
    
\node[title] at (-6,5.5) {Driving Forces};

\node[subtitle] at (-6,4.9) {Initial (Insufficient)};
\node[driving] (mgr) at (-6,4.1) {Managerial Interest};
\node[driving] (poc) at (-6,2.9) {IT PoC Started};
\node[driving] (emp) at (-6,1.7) {Employee Interest};

\draw[dasharrow, green!60!black] (mgr.east) -- (center.west|-mgr.east);
\draw[dasharrow, green!60!black] (poc.east) -- (center.west|-poc.east);
\draw[dasharrow, green!60!black] (emp.east) -- (center.west|-emp.east);
    
    \node[subtitle] at (-6,0.4) {With AI Task Force};
    \node[driving] (exec) at (-6,-0.5) {Executive Team Direction};
    \node[driving] (ai) at (-6,-2) {Technical AI Expertise};
    \node[driving] (bu) at (-6,-3.5) {Functional Experts};
    \node[driving] (leg) at (-6,-5) {Legal Representation};
    
    \draw[arrow, green!60!black] (exec.east) -- (center.west|-exec.east);
    \draw[arrow, green!60!black] (ai.east) -- (center.west|-ai.east);
    \draw[arrow, green!60!black] (bu.east) -- (center.west|-bu.east);
    \draw[arrow, green!60!black] (leg.east) -- (center.west|-leg.east);
    
    \node[title] at (6,5.5) {Restraining Forces};
    
    \node[restraining] (reg) at (6,3.5) {AI Act/Legal Framework};
    \node[restraining] (str) at (6,2) {Limited Strategic AI Alignment};
    \node[restraining] (it) at (6,0.5) {Barriers in IT Infrastructure};
    \node[restraining] (silo) at (6,-1) {Organizational Silos};
    
    \draw[arrow, red!60!black] (reg.west) -- (center.east|-reg.west);
    \draw[arrow, red!60!black] (str.west) -- (center.east|-str.west);
    \draw[arrow, red!60!black] (it.west) -- (center.east|-it.west);
    \draw[arrow, red!60!black] (silo.west) -- (center.east|-silo.west);

    \node[title] at (0,-5.9) {Legend};
    \draw[arrow, green!60!black] (-4,-7) -- (-2,-7);
    \node[align=left] at (0,-7) {Strong driving force};
    \draw[dasharrow, green!60!black] (-4,-8) -- (-2,-8);
    \node[align=left] at (0,-8) {Weak driving force};
    \draw[arrow, red!60!black] (-4,-7.5) -- (-2,-7.5);
    \node[align=right] at (0,-7.5) {Strong restraining force};
    
\end{tikzpicture}
\caption{Force Field Analysis of the AI Transformation}
\label{fig:forcefield}
\end{figure}

To understand the challenges X-FAIT addresses, we first define the restraining and driving forces of change, as illustrated in Figure~\ref{fig:forcefield}. We then describe how X-FAIT shifts the organizational equilibrium and accelerates transformation in the case company. 

\paragraph{\textbf{Restraining Forces of Change}}
Several key factors acted as barriers to AI-driven transformation at the case company:

\textit{Regulatory Constraints}: In a tightly regulated industry, compliance often took precedence over experimentation. Concerns about the upcoming EU AI Act\footnote{https://digital-strategy.ec.europa.eu/en/policies/regulatory-framework-ai}, set to take effect in 2025, further reinforced a cautious stance.

\textit{Limited Strategic AI Alignment}: 
AI initiatives had somewhat limited company-wide strategic alignment, leading to partially fragmented priorities and unclear evaluation metrics. While a shared IT department played a central role in enabling AI adoption, decision-making remained distributed across functions, making it difficult to align investments with broader business objectives.

\textit{Infrastructure and Process Barriers}: Existing IT structures and workflows were not optimized for AI due to: (1) legacy systems architecturally incompatible with contemporary AI requirements, (2) established procurement and implementation protocols ill-suited to iterative AI experimentation, (3) resource allocation frameworks that prioritized operational stability over innovation capacity, and (4) traditional development methodologies difficult to integrate with the adaptive, iterative nature of AI experimentation and implementation.

\textit{Organizational Fragmentation}: 
Siloed departments also created challenges for adoption by isolating expertise, limiting cross-functional learning, and leading to nonexistent or inconsistent AI strategies. Disconnected data and competing priorities also created challenges in resource allocation.

\paragraph{\textbf{Initial Driving Forces}}
Early interest from management and employees, along with initial proof-of-concept projects led by IT, signaled organizational readiness for AI. However, these efforts alone were not enough to overcome existing barriers due to limited organizational momentum, insufficient resource commitment, lack of cross-functional alignment, and the absence of a formalized implementation authority.

\paragraph{\textbf{Enhanced Driving Forces (With AI Task Force)}}
The introduction of the Cross-Functional AI Task Force (X-FAIT) accelerated organizational transformation by establishing a new equilibrium across key areas.

With executive sponsorship, X-FAIT generated momentum through three strategic mechanisms: aligning AI initiatives with corporate objectives, formalizing resource allocation to prioritize AI investments, and providing a clear mandate that reduced departmental resistance.

Instead of centralizing AI expertise within existing IT functions, X-FAIT included and embedded specialists from and directly into business and support functions. This approach enhanced implementation capability by enabling knowledge transfer across functions, refining technology evaluation for optimal use cases, adapting development and procurement processes for AI-specific needs, and applying risk assessment frameworks tailored to operational environments.

The integration of AI specialists within functional teams brought additional benefits. It ensured solutions were designed for practical business needs rather than general and potentially abstract technical possibilities, improved process optimization based on real-world constraints, facilitated user adoption through targeted change management, and measured impact using metrics relevant to each domain.

X-FAIT’s effectiveness was further reinforced by including legal and finance representatives, ensuring regulatory compliance and streamlined investment decisions. The task force also used existing talent by identifying employees with both domain expertise and AI aptitude. With executive support, these individuals became central to the organization’s AI risk management framework, forming a foundation for sustainable adoption.

In summary, the X-FAIT unified organizational support by converting isolated interests into coordinated momentum through formal structure and executive sponsorship. Furthermore, it integrated expertise by combining technical AI proficiency with domain knowledge and regulatory compliance, and focused resources by directing limited organizational resources toward strategically prioritized AI initiatives.

\section{Conclusion}

Our study demonstrates that while significant restraining forces often exist in AI transformation, a structured and strategically positioned task force can create counterbalancing driving forces of sufficient magnitude to overcome organizational inertia. By embedding risk management, strategic alignment, and cross-functional expertise within the X-FAIT framework, organizations can address multiple barriers simultaneously.

Through this Action Research engagement, we have seen early benefits in the case company, including improved cross-departmental collaboration, more structured risk assessment in AI deployment, and clearer alignment between AI initiatives and corporate strategy. By integrating existing theoretical models into a practical framework, this study provides both conceptual insights and actionable guidance for software-driven enterprises.

However, AI transformation is an evolving and iterative process, requiring continuous refinement of tools, methodologies, and governance structures. Future work should explore how AI transformation strategies can be adapted for organizations of different sizes and regulatory environments, as well as how long-term impact can be effectively measured. Advancing this field will require strong collaboration between industry and academia, ensuring that research-based insights translate into practical solutions that drive real-world AI adoption forward.


\section*{Acknowledgment}
ChatGPT and Claude.ai were utilized to improve the phrasing of some parts of the text, originally written by the authors.

\bibliographystyle{ACM-Reference-Format}
\bibliography{ref}

\end{document}